\documentstyle[12pt,epsfig]{article}
\def\slash{\hspace{-0.08in}}
\def\pslash{\hspace{-0.11in}}
\begin{document}
\baselineskip 20pt
\begin{center}
{\bf Meson-exchange $\pi N$ Models in
Three-Dimensional Bethe-Salpeter Formulation}\\
\vspace{0.5cm}
Cheng-Tsung Hung$^a$, Shin Nan Yang$^b$ and T.-S.H. Lee$^c$\linebreak
$^a$Chung-Hua Institute of Technology, Taipei, Taiwan 11522, ROC\\
$^b$Department of Physics, National Taiwan University, Taipei, Taiwan 10617, ROC\linebreak
$^c$Physics Division, Argonne National Laboratory, Argonne, Illinois 60439, U.S.A.
\end{center}

\begin{abstract}

The pion-nucleon scattering is investigated by using several
three-dimensional reduction schemes  of the Bethe-Salpeter
equation for a model  Lagrangian involving $\pi$, $N$,
$\Delta$, $\rho$, and $\sigma$ fields. It is found that all of the
resulting meson-exchange models can give similar good
descriptions of the $\pi N$ scattering data up to 400 MeV.
However they have significant differences in describing
the $\pi NN$ and $\pi N\Delta$ form factors and the $\pi N$
off-shell t-matrix elements.
We point out that these differences can be best distinguished by
investigating the near threshold pion production from nucleon-nucleon
collisions and pion photoproduction
on the nucleon. The consequences of using these models to
investigate various pion-nucleus reactions are also discussed.
\end{abstract}

\newpage

\section{Introduction}

Pion-nucleon interaction plays a fundamental role in determining the nuclear dynamics
involving pions. Despite very extensive investigations in the past two decades, several
outstanding problems remain to be solved. For example, an accurate description of pion
absorption by nuclei \cite{ashery,weyer,kamada,leea,satoa} is still not available and
hence the very extensive data for pion-nucleus reactions and pion productions from
relativistic heavy-ion collisions have not been understood satisfactorily. To make
progress, it is necessary to improve our theoretical description of the $\pi N$
off-shell amplitude which is the basic input to most of the existing nuclear
calculations at intermediate energies. The importance of the $\pi N$ off-shell t-matrix
in a dynamical description of pion photoproduction has also been demonstrated
\cite{yang85,lee,nbl90,satolee} in recent years.

Quantum Chromodynamics (QCD) is now commonly accepted as the
 fundamental theory of strong
interactions. However, due to the mathematical complexities, it is
not yet possible to predict $\pi N$ interactions directly from
QCD. On the other hand, models based on meson-exchange picture
\cite{paris,bonn} have been very successful in describing the $NN$
scattering. It is therefore reasonable to expect that the $\pi N$
dynamics at low and intermediate energies can also be described by the
same approach. Most of the recent attempts \cite{lee,pj,gs,hung,hung1,juelich}
in this direction were obtained by applying various
three-dimensional reductions of the Bethe-Salpeter equation for
$\pi N$ scattering.

As is well known \cite{klein}, the derivation of a three
dimensional formulation from the Bethe-Salpeter equation is not
unique. It is natural to ask whether the resulting off-shell
dynamics in the relevant kinematic regions depends strongly on the
choice of the reduction scheme. This question concerning the $NN$
models was investigated \cite{wj} quite extensively
in 1970's. No similar investigation for the $\pi N$ interactions
has been made so far. In this paper we report the progress we have
made on this question.

In section II, we specify the approximations that are used to
derive a class of three-dimensional $\pi N$ scattering equations
from the Bethe-Salpeter formulation. In section III, we define the
dynamical content of the resulting meson-exchange models.
The phenomenological aspects of the
models are described in section IV. The results and discussions
are presented in section V.

\section{Three-dimensional reduction of Bethe-Salpeter formulation}

To illustrate the derivations of three-dimensional equations
for $\pi N$ scattering from the Bethe-Salpeter formulation, it
is sufficient to consider a simple $\pi NN$ interaction
Lagrangian density
\begin{eqnarray}
 L_{int}(x) = \bar{\psi}(x) \Gamma_0 \psi(x)\phi(x)\, ,
\end{eqnarray}
where $\psi(x)$ and $\phi(x)$ denote respectively the nucleon and pion fields and
$\Gamma_0$ is a bare $\pi NN$ vertex, such as $\Gamma_0 = i g \gamma_5$ in the familiar
pseudo-scalar coupling. By using the standard method \cite{zuber}, it is straightforward
to derive from Eq. (1) the Bethe-Salpeter equation for $\pi N$ scattering and the
one-nucleon propagator. In momentum space, the resulting Bethe-Salpeter equation can be
written as
\begin{equation}
T(k^\prime,k;P) = B(k^\prime ,k;P) + \int d^4k^{\prime\prime}
B(k^\prime , k^{\prime\prime};P)G(k^{\prime\prime};P)T(k^{\prime\prime},k;P),
\label{BSeq}
\end{equation}
where $k$ and $P$ are respectively the relative and total momenta
defined by the nucleon momentum $p$ and pion momentum $q$
\begin{eqnarray}
P&=&p+q \, , \nonumber \\
k&=&\eta_\pi(y)p-\eta_N(y)q \, .\nonumber
\end{eqnarray}
Here $\eta_N(y)$ and $\eta_\pi(y)$ can be any function of a chosen parameter $y$ with
the condition
\begin{eqnarray}
\eta_\pi(y) + \eta_N(y) = 1 \, .
\end{eqnarray}
Obviously we have from  the above definitions that
\begin{eqnarray}
p&=&\eta_N(y)P+k \, , \nonumber \\
q&=&\eta_\pi (y)P-k \, .
\end{eqnarray}
In analogy to the nonrelativistic form, it is often to choose $\eta_N = m_N/(m_\pi+m_N)$
and $\eta_\pi=m_\pi/(m_\pi+m_N)$. The choice of  the $\eta'$s is irrelevant to the
derivation presented below in this section provided that Eq. (3) is satisfied.

Note that $T$ in Eq. (2) is the "amputated" invariant amplitude and is
related to the $\pi N$ S-matrix by
$S \propto \bar{u}T u$ with $u$ denoting the nucleon spinor.
The driving term
$B$ in Eq. (\ref{BSeq}) is the sum of all two-particle irreducible
amplitudes, and $G$ is the product of the pion propagator $D_{\pi}(q)$ and
the nucleon propagator $S_N(p)$.
In the low energy region, we  neglect the dressing of pion propagator
and simply set
\begin{eqnarray}
D_\pi(q) =\frac{1}{q^2-m_\pi^2+i\epsilon} \, ,
\end{eqnarray}
where $m_\pi$ is the physical pion mass.

The nucleon propagator
can be written as
\begin{eqnarray}
S_N(p) = \frac{1}{i p \hspace{-0.08in} / - m^0_N -\tilde\Sigma_N(p^2)+i\epsilon} \, ,
\end{eqnarray}
where $m^0_N$ is the bare nucleon mass and the nucleon self energy operator
$\tilde\Sigma_N$ is defined by
\begin{eqnarray}
\tilde\Sigma_N(p^2) =  \int d^4k \Gamma_0 G(k;p)\tilde\Gamma(k;p) \, . \label{nucl1}
\end{eqnarray}
The dressed vertex function $\tilde\Gamma$ on the right hand side of Eq.
(\ref{nucl1}) depends on the $\pi N$ Bethe-Salpeter amplitude
\begin{eqnarray}
\tilde\Gamma(k;P) = \Gamma_0 + \int d^4k^\prime \Gamma_0G(k^\prime;P) T(k^\prime,k;P) \,
. \label{nucl2}
\end{eqnarray}

It is only possible in practice
to consider the leading term of
 $B$ of Eq. (2). For the  Lagrangian Eq. (1) the
 leading term  consists of
the direct and crossed $N$ diagrams, as illustrated
 in Figs. 1(a) and 1(b)
\begin{eqnarray}
B (k,k';P) = B^{(a)}(k,k';P) + B^{(b)}(k,k';P) \, ,
\end{eqnarray}
where
\begin{eqnarray}
B^{(a)}(k,k';P) &=& \Gamma_0S_N(P) \Gamma_0, \\
 B^{(b)}(k,k';P) &=& \Gamma_0S_N(\bar P)\Gamma_0,
\end{eqnarray}
with {\it $\bar P =[ {\eta_N(y)-\eta_\pi(y)}]P+k+k'$.}

Equations (2)-(11) form a closed set of coupled equations
for determining
the dressed nucleon propagator of Eq. (6) and the
$\pi N$ Bethe-Salpeter amplitude of Eq. (2). It is important to
note here that this is a drastic simplification of the original
field theory problem defined by the Lagrangian Eq. (1).
However, it is still very difficult to
solve this highly nonlinear problem exactly.
For practical applications,
it is common to introduce further approximations.

The first step is to define the physical
nucleon mass by imposing the condition that
the dressed nucleon propagator
should have the limit {\it
\begin{eqnarray}
S_N(p) \rightarrow \frac{1}{i p \hspace{-0.08in} / - m_N+i\epsilon} \, ,
\end{eqnarray}}
as $p^2 \rightarrow m_N^2$ with $m_N$ being the physical nucleon mass.
This means that the self-energy in the nucleon propagator Eq. (6)
is constrained by the condition
\begin{eqnarray}
m^0_N + \tilde\Sigma_N(m_N^2) = m_N.
\end{eqnarray}
The next step is to assume that the $p$-dependence of the nucleon self-energy is weak
and we can use the condition Eq. (13) to set $m^0_N + \tilde\Sigma(p^2)\sim m^0_N
+\tilde\Sigma(m_N^2) = m_N$. This approximation greatly simplifies the nonlinearity of
the problem, since the full $\pi N$ propagator $G$ in Eqs. (2), (7) and (8) then takes
the following simple form
\begin{eqnarray}
G(k;P)=\frac{1}{i p \hspace{-0.08in} / - m_N+i\epsilon}
\,\frac{1}{q^2-m^2_\pi+i\epsilon}.
\end{eqnarray}
To be consistent, the driving terms Eqs. (10) and (11) are also
evaluated by using the simple nucleon propagator of the form of
Eq. (12).

The next commonly used approximation is to reduce the
dimensionality of the above integral equations from four to three.
In addition to simplifying the numerical task, this is also
motivated by the consideration that the above covariant
formulation is not consistent with most of the existing nuclear
calculations based on the three-dimensional Schroedinger
formulation.

The procedure for reducing the dimensionality of the above
equations is to replace the propagator $G$ of Eq. (14), by a
propagator $\hat{G}_0$ which contains a $\delta$-function
constraint on the momentum variables. In the low energy region,
this new propagator must be chosen such that the resulting
scattering amplitude has a correct $\pi N$ elastic cut from
$(m_\pi+m_N)^2$ to $\infty$ in the complex $s$-plane, as required
by the unitarity condition. It is well known (for example, see
Ref. \cite{klein}) that the choice of such a $\hat{G}_0$ is rather
arbitrary. In this work, we focus on a class of three dimensional
equations which can be obtained by choosing the following form
\begin{eqnarray}
\hat G_0(k;P) &= & \frac{1}{(2\pi)^3}\int \frac{ds'}{ s-s'+i\epsilon} f(s,s') [\alpha
(s,s') P \hspace{-0.10in} / +k \hspace{-0.08in} / +m_N]  \nonumber  \\
 & \times& \delta^{(+)}([\eta_N(s')P' + k]^2
- m_{N}^2)  \delta^{(+)}
([\eta_\pi (s')P'- k]^2 -  m_{\pi}^2).  \label{2GH}
\end{eqnarray}
In the above equation, $s=P^2$ is the invariant mass of the $\pi N$ system, and
$P'=\sqrt{\frac{s'}{s}}P$ defines the "offshellness" of the intermediate states. The
superscript (+) associated with $\delta$-functions means that only the positive energy
part is kept in defining the  nucleon propagator. The relative momentum $k$ in the
$\delta$-functions is defined by setting $y=s$ in $\eta's$, i.e., $k  =
\eta_\pi(s)p-\eta_N(s)q$. To have a correct $\pi N$ elastic cut, the arbitrary functions
$f(s,s')$ and $\alpha(s,s')$ must satisfy the conditions
\begin{eqnarray}
 f(s,s)  &=&  1, \\
\alpha(s,s)&=&\eta_N(s).
\end{eqnarray}
It is easy to verify that
for $(m_\pi+m_N)^2 \leq s \leq \infty$,  Eqs. (15)-(17)
give the correct discontinuity of the propagator $\hat{G}_0$
\begin{eqnarray}
Disc[\hat{G}_0(k;P)] & =&\frac{-i}{(2\pi)^2}
(\eta_N(s) P \hspace{-0.07cm}\slash / + k \slash /+m_N) \delta^{(+)}([\eta_N(s) P + k]^2-m_N^2) \nonumber \\
       &\times &
          \delta^{(+)}([\eta_\pi (s) P -k]^2 - m_\pi^2).
\end{eqnarray}

Several three dimensional formulations developed in the literature
can be derived from using Eqs. (15)-(17). These are given by
Blankenbecler and Sugar ($BbS$) \cite{bs}, Kadyshevsky ($Kady$) \cite{kady},
Thompson ($Thomp$) \cite{thom}, and Cooper and Jennings ($CJ$) \cite{cj}.
In Table 1, we list their
choices of the functions {\it $f(s,s')$} and {\it $\alpha(s,s')$}.
All schemes set
$\eta_N(s)=\varepsilon_N(s)/(\varepsilon_N(s)+\varepsilon_\pi(s))$
and $\eta_\pi(s) =
\varepsilon_\pi(s)/(\varepsilon_N(s)+\varepsilon_\pi(s)),$ where
$\varepsilon_N(s)=(s+m_N^2-m_\pi^2)/2\sqrt{s}$ and
$\varepsilon_\pi(s)=(s-m_N^2+m_\pi^2)/2\sqrt{s}$ are the
center of mass (CM) energies of nucleon and pion, respectively.


In the rest of the paper, we will present the formulation in the CM frame.
In this frame, we
have {\it $P=( \sqrt{s},\vec{0})$} for the total momentum,
$\vec{p} =\vec{k}$ and $\vec{q}=-\vec{k}$.
 The integral over $s'$ in  Eq. (\ref{2GH})
can then be carried out to yield
\begin{equation}
\hat G_0(\vec{k};\sqrt{s})
=\frac{1}{(2\pi)^3} \frac{\delta(k_0-\hat{\eta}(s_{\vec{k}},
\vec{k}))} {\sqrt{s}-\sqrt{s_{\vec{k}}}+i\epsilon} \frac{2\sqrt{s_{\vec{k}}}} {\sqrt{s}+
\sqrt{s_{\vec{k}}}} f(s,s_{\vec{k}}) \frac{\alpha(s,s_{\vec{k}}) \gamma_0 \sqrt{s} +k
\hspace{-0.08in} / + m_N} {2E_N(\vec{k})2E_\pi(\vec{k})},
\end{equation}
where  $E_N(\vec{k})=(\vec{k}^2+m_N^2)^{1/2}$
 and  $E_\pi(\vec{k})=(\vec{k}^2+m_\pi^2)^{1/2}$ are
the nucleon and pion energies, and we have defined
\begin{eqnarray}
 \sqrt{ s_{\vec{k}} }  &=&  E_N(\vec{k})+E_\pi(\vec{k}) \nonumber \\
\hat{\eta}(s,\vec{k})  &=&
 \frac 12 [\sqrt{s}+ E_N(\vec{k})-E_\pi(\vec k) -
2\eta_N(s) \sqrt{s}]. \nonumber
\end{eqnarray}
Replacing $G$ by $\hat{G}_0$ in Eq. (2) and performing the
integration over the
time component $k''_0$, we then obtain a three-dimensional
scattering equation of the following
form
\begin{equation}
t(\vec{k'},\vec{k};\sqrt{s}) = v(\vec{k'},\vec{k};\sqrt{s})
+ \int d \vec{k''}v(\vec{k'},\vec{k''};\sqrt{s})
g(\vec k'';\sqrt{s}) t(\vec{k''},\vec{k};\sqrt{s}),
\end{equation}
where
\begin{eqnarray}
t(\vec{k'},\vec{k};\sqrt{s})
& = &\int dk'_0 dk_0\delta(k'_0-\hat{\eta}') T(k',k;\sqrt{s})
\delta(k_0-\hat{\eta}),   \\
v(\vec{k'},\vec{k};\sqrt{s})
& = &\int dk'_0 dk_0\delta(k'_0-\hat{\eta}') B(k',k;\sqrt{s})
\delta(k_0-\hat{\eta}),  \\
g(\vec k;\sqrt{s}) & = & \int dk_0 \hat G_0(k;\sqrt{s}), \label{g1}
\end{eqnarray}
with $\hat{\eta}'=  \hat{\eta}(s_{\vec{k'}}, \vec{k'})$ and $\hat{\eta}=
\hat{\eta}(s_{\vec{k}}, \vec{k})$.\\

Substituting
the $\alpha's$ and $f's$ listed in Table 1 into (23), we
find \cite{hung} that the propagator of the three-dimensional
scattering equation Eq. (20) for each reduction scheme is
\begin{enumerate}
\item {\em Cooper-Jennings propagator}
\begin{eqnarray*}
 g(\vec{k};\sqrt{s})  =  \frac{1}{(2\pi)^3}\,
 \frac{1}{\sqrt{s}-\sqrt{s_{\vec{k}}}+i\epsilon} \,\frac{2\sqrt{s_{\vec{k}}}} {\sqrt{s}+
 \sqrt{s_{\vec{k}}}}\, \frac{\sqrt{ss_{\vec{k}}}}{ss_{\vec{k}}-(m_N^2-m_\pi^2)^2}
 \left[ \gamma_0 \varepsilon_N(s)-\vec{\gamma} \cdot \vec{k} +m_N \right].
\end{eqnarray*}

\item {\em Blankenbecler-Sugar propagator}
 \begin{eqnarray*}
  g(\vec{k};\sqrt{s})   =  \frac{1}{(2\pi)^3}\,
  \frac{1}{\sqrt{s}-\sqrt{s_{\vec{k}}}+i\epsilon}\, \frac{2\sqrt{s_{\vec{k}}}} {\sqrt{s}+
  \sqrt{s_{\vec{k}}}}\, \frac{1}{4E_N(\vec{k}) E_\pi(\vec{k})} \left[ \gamma_0
  E_N(\vec{k})-\vec{\gamma} \cdot \vec{k} +m_N \right].
 \end{eqnarray*}

\item {\em Thompson propagator}
 \begin{eqnarray*}
  g(\vec{k};\sqrt{s})  =  \frac{1}{(2\pi)^3}\,
  \frac{1}{\sqrt{s}-\sqrt{s_{\vec{k}}}+i\epsilon} \sqrt{\frac{s_{\vec{k}}}{s}}\,
  \frac{1}{4E_N(\vec{k}) E_\pi(\vec{k})}\,
  \left[ \gamma_0 \varepsilon_N(s)-\vec{\gamma} \cdot \vec{k} +m_N \right].
 \end{eqnarray*}

\item {\em Kadyshevsky propagator}
 \begin{eqnarray*}
  g(\vec{k};\sqrt{s})  =  \frac{1}{(2\pi)^3}\,
  \frac{1}{\sqrt{s}-\sqrt{s_{\vec{k}}}+i\epsilon}\, \frac{1}{4E_N(\vec{k}) E_\pi(\vec{k})}
  \left[ \gamma_0 E_N(\vec{k})-\vec{\gamma} \cdot \vec{k} +m_N \right].
 \end{eqnarray*}
\end{enumerate}

If we  consistently replace $G$ by $\hat{G}_0$ in evaluating
Eqs.  (\ref{nucl1}-\ref{nucl2}), we then also obtain  a
 numerically much simpler
three-dimensional form $\Sigma_N$ for
the nucleon self energy $\tilde{\Sigma}_N$ and $\Gamma$ for the
dressed vertex function $\tilde{\Gamma}$.
The resulting equations in the CM frame are
\begin{eqnarray}
\Sigma_N(\sqrt{s})
&=& \int d\vec{k} \Gamma_0 g(\vec{k};\sqrt{s})
\Gamma(\vec{k};\sqrt{s}), \\
\Gamma(\vec{k};\sqrt{s})
&=&\Gamma_0 + \int d\vec{k'} \Gamma_0 g(\vec{k}^{\, '};\sqrt{s})
t(\vec{k}^{\, '},\vec{k};\sqrt{s}).
\end{eqnarray}
Accordingly, the nucleon pole condition Eq. (13) becomes
\begin{eqnarray}
m^0_N + \Sigma_N(m_N) = m_N
\end{eqnarray}
This completes the derivations of the
three-dimensional formulations considered in this work.

\section{ Model Lagrangian and the $\pi N$ potentials}

 To define the $\pi N$ potential by using Eq. (22), we
assume that the driving term $B(k^\prime,k;\sqrt{s})$
is the sum of all tree diagrams calculated from the
following interaction Lagrangian
\begin{eqnarray}
{\cal L}_I & = & \frac{f^{(0)}_{\pi NN}}{m_\pi} \bar N \gamma_5 \gamma_\mu
 \vec{\tau} \cdot \partial^\mu \vec{\pi} N
    -g^{(s)}_{\sigma\pi\pi}m_{\pi} \sigma (\vec{\pi} \cdot \vec{\pi})
    -\frac{g^{(v)}_{\sigma\pi\pi}}{2m_\pi}\sigma \partial^{\mu}\vec{\pi}\cdot\partial_{\mu}\vec{\pi}
     \nonumber \\
    & &  -g_{\sigma NN}\bar{N}\sigma N
   -g_{\rho NN} \bar{N} \{ \gamma_\mu \vec{\rho}\,{}^\mu +
\frac{\kappa_V^\rho}{4m_N} \sigma_{\mu\nu} (\partial ^\mu \vec{\rho}\,{}^\nu -
\partial^\nu \vec{\rho}\,{}^\mu) \} \cdot \frac{1}{2}\vec{\tau} N  \nonumber \\
& & -g_{\rho\pi\pi} \vec{\rho}\,{}^{\mu} \cdot (\vec{\pi} \times \partial_{\mu}
 \vec{\pi}) -
\frac{g_{\rho\pi\pi}}{4m_{\rho}^2}(\delta - 1)(\partial^\mu \vec{\rho}\,{}^\nu -
\partial^\nu \vec{\rho}\,{}^\mu) \cdot (\partial_\mu \vec{\pi} \times
\partial_\nu \vec{\pi})  \nonumber \\
& & + \{\frac{g^{(0)}_{\pi N\Delta}}{m_\pi}  \bar\Delta_\mu [g^{\mu\nu}
-(Z+\frac{1}{2})\gamma^\mu\gamma^\nu] \vec{T}_{\Delta N}N \cdot
 \partial_{\nu} \vec \pi + h.c.\}, \label{largrang}
\end{eqnarray}
where $\Delta_{\mu}$ is the Rarita-Schwinger field operator for the $\Delta$, $\vec
T_{\Delta N}$ is the isospin transition operator between the nucleon and the $\Delta$.
The notations of Bjorken-Drell \cite{bd} are used in Eq. (\ref{largrang}) to describe
the field operators for the nucleon $N$, the pion $\vec{\pi}$, the rho meson
$\vec{\rho}$, and a fictitious scalar meson $\sigma$.  For $\sigma\pi\pi$ coupling, a
mixture of the scalar and vector couplings is introduced to simulate the broad width of
the S-wave correlated two-pion exchange mechanism \cite{hung1,juelich}. As illustrated
in Fig. 1, the resulting driving term consists of the direct and crossed $N$ and
$\Delta$ terms, and the t-channel $\sigma$- and $\rho$-exchange terms.

To write down the resulting matrix elements of the $\pi N$
potential, defined by Eq. (22), we introduce the following notations:
$q=(E_\pi(k), \vec{k})$ is the four-momentum
for the pion and $ p = (E_N(k), -\vec{k})$ for the nucleon.
The nucleon helicity is denoted as $\lambda$. We then have (isospin factors are
suppressed here)
\begin{eqnarray}
v(\vec{k}^{\,\prime},\vec{k};\sqrt{s})=
\sum_{\alpha=a,..f} V^{(\alpha)}(p',q';p,q).
\end{eqnarray}
The diagrams (a)-(b) of Fig. 1 give
\begin{equation}
V^{(a)}(p',q';p,q) = (\frac{f^{(0)}_{\pi NN}}{m_\pi})^2 \gamma_5
q'\pslash /
 \frac{p\slash / +
q\slash / + m_N} {(p+q)^2-m_N^2} \gamma_5 q\slash /, \\
\end{equation}

\begin{equation}
 V^{(b)}(p',q';p,q) = (\frac{f^{(0)}_{\pi NN}}{m_\pi})^2 \gamma_5  q\slash /
\frac{p'\pslash / - q\slash / + m_N} {(p'-q)^2-m_N^2} \gamma_5
q'\pslash / .
\\
\end{equation}
The $\sigma$-exchange diagram Fig. 1(c) has
a  component from the
scalar coupling and a component from the vector coupling
\begin{eqnarray}
V^{(c-s)}(p',q';p,q) & = & g_{\sigma NN}g_{\sigma\pi\pi}^{(s)}{m_\pi}  \frac{1}
{(p-p')^2-m_\sigma^2} , \\
 V^{(c-v)}(p',q';p,q) & = & \frac{g_{\sigma NN}g_{\sigma\pi\pi}^{(v)}}{2m_\pi}
\frac{q'\cdot q} {(p-p')^2-m_\sigma^2},
\end{eqnarray}
while the $\rho-$exchange diagram of Fig. 1(d) gives
\begin{eqnarray}
 V^{(d)}(p',q';p,q)   =   -g_{\rho NN}g_{\rho \pi\pi}  \frac{B_1 q\slash / +
 B_2 q'\pslash / + B_3 + B_4} {(p-p')^2-m_\rho^2},
\end{eqnarray}
with
\begin{eqnarray}
&&B_1   =   (1+ \kappa_V^\rho) (1+\frac{\delta-1}{4m_\rho^2})(p-p') \cdot q',
\nonumber \\
&&B_2  =   -(1+\kappa_V^\rho) \frac{\delta-1}{4m_\rho^2} (p-p') \cdot q,
\nonumber \\
&&B_3   =  - \frac{\kappa_V^\rho}{2m_N}[1+\frac{\delta-1}{4m_\rho^2}(p-p') \cdot
q'] (p+p') \cdot q, \nonumber \\
&&B_4  =   \frac{\kappa_V^\rho}{2m_N}\frac{\delta-1}{4m_\rho^2}[(p-p') \cdot q]
[(p+p') \cdot q'].
\end{eqnarray}
The contributions from the $\Delta$ excitations are  depicted in
diagrams of Figs. 1(e) and 1(f)
\begin{eqnarray}
 V^{(e)}(p',q';p,q) &  =  &- (\frac{g^{(0)}_{\pi N\Delta}}{m_\pi})^2
[g^{\mu\mu'}-(Z+\frac{1}{2}) \gamma^{\mu'} \gamma^{\mu}] \nonumber \\
 & & \times
\frac{2m_\Delta q'_{\mu'}\Lambda^{\mu\nu}(p+q)q_{\nu'}}{(p+q)^2-m_\Delta^2}
[g^{\nu'\nu}-(Z+\frac{1}{2}) \gamma^\nu \gamma^{\nu'}],
\end{eqnarray}

\begin{eqnarray}
 V^{(f)}(p',q';p,q) &  =  &- (\frac{g_{\pi N\Delta}}{m_\pi})^2
[g^{\mu\mu'}-(Z+\frac{1}{2}) \gamma^{\mu'} \gamma^{\mu}] \nonumber \\
 & & \times
\frac{2m_\Delta q_{\mu'}\Lambda_{\mu\nu}(p-q')q'_{\nu'}}{(p-q')^2-m_\Delta^2}
[g^{\nu'\nu}-(Z+\frac{1}{2}) \gamma^\nu \gamma^{\nu'}],
\end{eqnarray}
where $\Lambda_{\mu\nu}$ is
\begin{equation}
\Lambda_{\mu\nu}(P_\Delta)= \frac{{P\pslash /}_\Delta + m_\Delta}{2m_\Delta}[
g_{\mu\nu} - \frac{1}{3} \gamma_\mu \gamma_\nu
 - \frac{2 P_{\Delta\mu}P_{\Delta\nu}} { 3 m_\Delta^2} + \frac{P_{\Delta\mu}\gamma_\nu
 -P_{\Delta\nu}\gamma_\mu}{3m_\Delta}].\\
\end{equation}

The partial-wave decomposition of these potential matrix elements was discussed in
detail in Ref. \cite{hung}.

 \section{Renormalizations in $P_{11}$ channel}

Because of the appearance of one-particle intermediate state in
of Fig. 1(a), the $\pi N$ scattering amplitude, defined by
Eq. (20), in $P_{11}$ channel can be decomposed
into a sum of pole  and non-pole (background) terms.
 In the operator form, the $P_{11}$ amplitude can be
written as
\begin{eqnarray}
t(E)= t^{bg}(E) +
\frac{\Gamma^\dagger(E^*)|N_0><N_0| \Gamma(E)}{E
-m^0_N - \Sigma_N(E)},
\end{eqnarray}
where $|N_0>$ is the bare one-nucleon state and
\begin{eqnarray}
t^{bg}(E)&=& v^{bg}(E) + v^{bg}(E) g(E)t^{bg}(E), \\ \Gamma(E)&=& \Gamma_0[ 1 + g(E)
t^{bg}(E)], \label{Gamma}\\ \Sigma_N(E) &=& <N_0|\Gamma_0 g(E)\Gamma^\dagger(E^*)|N_0>.
\end{eqnarray}
In the above equations, $E=\sqrt{s}+i\epsilon$ and $\Gamma_0 $ denotes the bare $N_0
\rightarrow \pi N$ vertex in Fig. 1(a). $t^{bg}$ is  due to the background potential
$v^{bg}$ which is the sum of contributions (b), (c), (d), and (f) of Fig. 1. $\Gamma$ is
the dressed $\pi NN$ vertex. We follow the procedure of Afnan and his collaborators
\cite{afnan} to constrain the fit of $P_{11}$ phase shifts by imposing the nucleon pole
condition Eq. (26). This also leads to a condition which relates the bare coupling
constant $f_{\pi NN}^{(0)}$ to the empirical $\pi NN$ coupling constant.

As $E \rightarrow m_N$, the self-energy $\Sigma_N(E)$
 can be expressed as
\begin{eqnarray}
\Sigma_N(E)=\Sigma_N(m_N)+(E-m_N)\Sigma_1(m_N)
+ \cdot\cdot\cdot
\end{eqnarray}
where
\begin{equation}
\Sigma_1(m_N)= \left. \frac{\partial \Sigma_N(E)}{\partial E}  \right
|_{E=m_N},
\end{equation}

 The above relations lead to a
renormalization of the $\pi NN$ coupling constant. The
renormalized coupling constant $f_{\pi NN}$ is related to the bare
coupling constant $f_{\pi NN}^{(0)}$
by
\begin{eqnarray}
f_{\pi NN} = f_{\pi NN}^{(0)} [ 1 + g(m_N) t^{bg}(m_N)]Z_2^{1/2}.
\end{eqnarray}
where the nucleon wave function renormalization constant is given by
\begin{eqnarray}
Z_2^{-1}=1+\Sigma_1(m_N).
\end{eqnarray}
The renormalized coupling constant is identified with the empirical value
 $g^2_{\pi NN}/4\pi=(2m_N/m_\pi)^2(f_{\pi NN}^2/4\pi)=
14.3$. Equations for $P_{33}$
channel can also be written in the form of Eqs. (39)-(43) with $N$
replaced by $\Delta$.

\section{The parameters and the Fitting Procedures}

To complete the model we need to introduce form factors to regularize
the potential matrix elements defined by Eqs. (28)-(38).
In this work we follow Pearce and Jennings \cite{pj}
and associate each external leg of the
potential matrix elements  with a form factor of the form
\begin{equation}
 F(\Lambda, p)=\left[\frac{n\Lambda^4}{n\Lambda^4+(p^2-m^2)^2} \right]^{n},  \label{ff}
\end{equation}
where $p=(p_0, \vec{p})$ with $p_0=(m_N^2+p_E^2)^{1/2}$ defined by the on-shell momentum
$p_E$ of the incident energy. It is interesting to note that as $n \rightarrow \infty$,
$F(\Lambda , p)$ approaches to a Gaussian form.

The parameters which are allowed to vary in fitting the empirical
 phase shifts are:
$(g_{\sigma NN}g^{(s)}_{\sigma\pi\pi}),\,
(g_{\sigma NN}g^{(v)}_{\sigma\pi\pi}),\,
(g_{\rho NN}g_{\rho\pi\pi})$, and $\delta$ for the t-channel $\sigma$ and $\rho$ exchanges,
$m_{\Delta}^{(0)},\, g^{(0)}_{\pi N\Delta},\, Z$ for the $\Delta$ mechanisms, and the
cut-off parameters $\Lambda$'s of the form factors of Eq. (\ref{ff}). In the crossed $N$
diagram, the physical $\pi NN$ coupling constant is used. For the crossed $\Delta$
diagram, the situation is not so clear since the determination of the "physical" $\pi
N\Delta$ coupling constant depends on the nonresonant contribution in the $P_{33}$
channel. In principle, it can be determined by carrying out a renormalization procedure
similar to that used for the nucleon. However, it is a much more difficult numerical
task. The complication is due to the fact that the $\Delta$ pole is complex. As in Ref.
\cite{lee,pj,gs}, such a renormalization for the $\Delta$ is not carried out in this work
and we simply allow the coupling constant used in the crossed $\Delta$ diagram to also
vary in the fit to the data. This coupling constant is denoted as $g_{\pi N\Delta}$.\\

\section{Results and Discussions}

We first consider the models using rank $n=2$ form factor defined by Eq. (\ref{ff}). The
constructed models are called $C2$, $B2$, $T2$, and $K2$ for the Cooper-Jennings,
 Blankenbecler-Sugar, Thompson, and Kadyshevsky reduction schemes,
respectively. For each model, we adjust the parameters
described in the previous section to fit the data of $\pi N$ scattering phase
shifts \cite{VPI97}. The results for the $K2$ model is shown in Fig. 2.
We see that the data can be described very well. The results of other
three models are very similar in all channels except in
the $P_{11}$ channel. This is illustrated in Fig. 3. The difficulty in getting the same
fit to this channel is mainly due to the nucleon
renormalization conditions Eqs. (26) and (44).
This difficulty is well known in the literature. Our results for the K2, B2 and T2 are
acceptable. We, however, are not able to improve the result for C2 unless we ignore the
fit to other channels.

The resulting parameters of the constructed four models are listed in Table 2. We first
notice that the bare $\pi NN$ coupling constant
$g^{(0)}_{\pi NN}=(2m_N/m_\pi)f_{\pi NN}^{(0)}$ is considerably smaller than the
physical value $g_{\pi NN}$ in all models. This large vertex renormalization is closely
related to an about 150 MeV mass shift between the bare mass $m^{(0)}_N$ and $m_N$, as
seen in the first two rows of Table 2. The determined physical coupling constant $g_{\pi
N \Delta}$ for the crossed $\Delta$ term, Fig. 1(f), is also significantly larger than the
bare coupling constant $g^{(0)}_{\pi N \Delta}$. The large difference between the bare
mass $m^{(0)}_\Delta \sim 1400$ MeV and the resonance position $m_\Delta=1232$ MeV seems
to be a common feature of the constructed models.

The parameters associated with the $\rho$-exchange are comparable to that of
other meson-exchange $\pi N$ models. The $\sigma-$exchange turns out
to be not important in the fit. If we set the coupling constant
$g_{\sigma NN}g_{\sigma \pi\pi}^{(v)}$ of all models to
zero, the resulting phase shifts are not changed much.
This is consistent with Ref. \cite{satolee} in which the fit was achieved
without including a $\sigma$-exchange mechanism.

It is also interesting to note that the fit to the data seems to favor a soft $\pi NN$
form factor with $\Lambda_\pi \le 700$ MeV for the models $C2$, $B2$, and $T2$. The
value $\Lambda_\pi \sim 860$ MeV for the model K2 is also not too hard compared with the
range used in defining nucleon-nucleon potential and
consistent with previous findings \cite{satolee,pj}.

An essential phenomenology in constructing the
meson-exchange models is the use of form
factors to regularize the potential. To develop theoretical interpretations of the
determined parameters listed in Table 2, it is important to investigate how the models
depend on the parameterization of the form factors. For this we also consider models
with very high rank form factors defined by Eq. (\ref{ff}) with $n=10$.
As discussed in Ref. \cite{pj},
this very high rank form is close to the Gaussian form. We find that a fit
comparable to that shown in Figs. 2 and 3 can also be obtained with this parameterization
of form factors. There are some significant, though not very large, changes in the
resulting parameters. This is illustrated in Table 3 in which the parameters from using
$n=2$ (T2 and K2) and $n=10$ (T10 and K10) form factors are compared.

The constructed four models can be considered approximately
phase-shift equivalent. We
therefore can examine how the  resulting $\pi N$ off-shell
 dynamics depends on the chosen three-dimensional reduction.
The $\pi N$ off-shell amplitudes are
needed to study nuclear dynamics involving pions. To be
specific, let us first discuss how the constructed models can be used to investigate the
near threshold pion production from nucleon-nucleon collisions. The most important
leading mechanism of this reaction is
that a pion is emitted by
one of the nucleons and then scattered from the
second nucleon. The matrix element of this rescattering
mechanism can be predicted by using the dressed $\pi NN$ form factor and
the half-off-shell t-matrix.
The predicted $\pi NN$ form factors for the near threshold kinematics,
  $E = m_N$, are compared in Fig. 4. In Fig. 5, we compare the
half-off-shell t-matrix elements in the  most relevant
$S_{11}$ and $S_{31}$ channels at at pion lab. energy $1 \, MeV$ above threshold. We see that there are rather significant
differences between the considered reduction schemes
 at $k \geq 500$ MeV/c which is close to the momentum of the exchanged pion
at the production threshold.
Consequently,  a study of near threshold pion production
from NN collisions could distinguish the considered four different reduction schemes.

We next discuss the reactions at the $\Delta$ excitation region. In Figs. 6 and 7, we
show the predicted dressed $\pi N \rightarrow \Delta$ form factor $\Gamma_{\pi N\Delta}$,
defined analogously to $\Gamma_{\pi NN}$ of Eq. (\ref{Gamma}), and the half-off-shell
t-matrix at the $\Delta$ resonance energy. These quantities are the input to the
investigations of the $\Delta$ excitation in pion photoproduction
\cite{yang85,nbl90,satolee}. The results shown in Figs. 6 and 7 suggest that the
considered reduction schemes can also be distinguished by investigating the pion
photoproduction reactions. This however requires a consistent derivation of the
photoproduction formulation for each reduction scheme, and is beyond the scope of this
work.

The differences shown in  Figs. 6 and 7  can also have important consequences in
determining pion-nucleus reactions in the $\Delta$ region.
For instance, the constructed
four models will give rather different predictions of pion double-charge reactions which
are dominated by two sequential off-shell $\pi N$ single-charge exchange scattering.
They can also be distinguished by investigating pion absorption which is induced by the
dressed $\pi N \rightarrow \Delta$ vertex, Fig. 6, followed by a $N\Delta \rightarrow NN$
transition.

In summary, we have shown that the $\pi N$ scattering data up to 400 MeV can be equally
well described by four reduction schemes of Bethe-Salpeter equation.
The resulting meson-exchange models yield rather different off-shell dynamics. With the
high quality data obtained in recent years, they can be best distinguished by
investigating pion productions from NN collisions and pion photoproductions. Their
differences in describing pion-nucleus reactions are also expected to be significant.
Our effort in these directions will be published elsewhere. \\

\centerline{\bf Acknowledgment}
We thank Drs. B. Pearce and C. Sch\"utz for useful
communications concerning their works. C.T.H. also wishes to thank Mr. Guan-yeu
Chen for checking part of the program.   
This work is supported in part by the National Science Council of ROC under
grant No. NSC82-0208-M002-17
and the U.S. Department of Energy, Nuclear
Physics Division, under contract No.W-31-109-ENG-38, and also in part by
the U.S.-Taiwan National Science Council Cooperative Science Program
grant No. INT-9021617.

\newpage

\vspace{1.0cm}
\begin{table}[htbp]
\begin{center}
\large{
\begin{tabular}{||c|c|c|c|c||}
\hline
  &{\em BbS} & {\em Kady} & {\em Thomp} & {\em CJ}  \\
 \hline
&&&&\\
$\alpha(s,s')$&  $\eta_N(s')\sqrt{\frac{s'}{s}}$ & $\eta_N(s')\sqrt{\frac{s'}{s}}$
 & $\eta_N(s)$ &$\eta_N(s)$\\
&&&& \\
 \hline
&&&&\\
 $f(s,s')$ &1   &$\frac{\sqrt{s}+\sqrt{s'}}{2\sqrt{s'}}$
 &$\frac{\sqrt{s}+\sqrt{s'}}{2\sqrt{s}}$
 &$\frac{4\sqrt{ss'}\varepsilon_N(s')\varepsilon_\pi(s')}
 {ss'-(m_N^2-m_\pi^2)^2}$ \\
&&&& \\
 \hline
 \end{tabular}
 }
\caption{The functions $\alpha(s,s')$ and $f(s,s')$ of Eq. (15),
  chosen for the four considered reduction schemes, i.e., Blankenbecler and Sugar ($BbS$) \cite{bs}, Kadyshevsky ($Kady$) \cite{kady},
Thompson ($Thomp$) \cite{thom}, and Cooper and Jennings ($CJ$) \cite{cj}.}
 \end{center}
\end{table}
 \vspace{1.0cm}

\begin{table}
\begin{center}
\begin{small}
\begin{tabular}{ccccc}
{\normalsize Parameter}  &{\normalsize C2} & {\normalsize  B2} & {\normalsize T2} &{\normalsize K2}   \\ \hline
$m_N$                          &939   & 939  & 939   & 939     \\
$m_N^{(0)}$                    &1090   & 1072   &1071   &1116       \\
$m_\pi$                        &137     & 137     & 137     & 137       \\
$m_\Delta$                     &1232    & 1232   &1232    & 1232.    \\
$m_\Delta^{(0)}$                   &1415 &1412   &1410   &1461     \\
$m_\rho$                       &770     &770     &770     &770       \\
$m_\sigma$                     &654    &662     &654     &654.      \\
$g_{\pi NN}^2/4\pi$                &14.3     &14.3      &14.3      & 14.3       \\
$g_{\pi NN}^{(0)^2}/4\pi$              &3.82   &6.28    &5.49    &6.08     \\
$g_{\sigma NN}g_{\sigma \pi\pi}^{(s)}/4\pi$    &-0.49 &-0.37   &-0.50   &-0.39     \\
$g_{\sigma NN}g_{\sigma \pi\pi}^{(v)}/4\pi$    &33.20  &-1.53  &-1.40   &-1.40     \\
$g_{\rho NN}g_{\rho\pi\pi}/4\pi$           &2.54   &2.87   &2.87    &2.90     \\
$\delta$                       &1.02   & 1.05  &1.06  &1.10     \\
$g_{\pi N\Delta}^2/4\pi$               &0.41  & 0.31  &0.29   &0.34      \\
$g_{\pi N\Delta}^{(0)^2}/4\pi$             &0.14   & 0.17  &0.17   &0.18      \\
$Z$                        &-0.14 &-0.036 &-0.075 &-0.029\\
$\kappa_V^\rho$                    &1.00   &1.00    &1.19   &1.55     \\
\\
$\Lambda_N$                    & 1227  &1383    &1321   &1239     \\
$\Lambda_\sigma$                   &  417 &704    & 681   &648       \\
$\Lambda_\rho$                     & 1521 &1700    &1637    &1548     \\
$\Lambda_\Delta$                   & 1026 & 1555   &1542   &1429     \\
$\Lambda_\pi$                      & 674   & 690    &666    &859       \\
\hline
\end{tabular}
\end{small}
\end{center}
\caption{The parameters of the constructed meson-exchange
   models, defined by Eqs. (29)-(36), are compared.
   The form factor Eq. (\ref{ff}) with $n=2$ is used.
   The models are constructed by using the three dimensional
   reduction schemes of Cooper and Jennings (C2),
   Blankenbecler and Sugar (B2), Thompson (T2), and Kadyshevsky (K2).}
\end{table}

\begin{table}[thbp]
\begin{center}
\begin{small}
\begin{tabular}{ccccc}
{\normalsize Parameter}  &{\normalsize T10} & {\normalsize  T2} & {\normalsize K10} &{\normalsize K2}   \\ \hline
$m_N$                          & 939    & 939   & 939   & 939     \\
$m_N^{(0)}$                    & 1065  &1071    &1073    &1116      \\
$m_\pi$                        & 137    & 137     & 137    & 137       \\
$m_\Delta$                     &1232   &1232    & 1232   & 1232     \\
$m_\Delta^{(0)}$                   &1407   &1410    &1420   &1461     \\
$m_\rho$                       &770     &770   & 770    &770       \\
$m_\sigma$                     &654     &654     & 654    &654       \\
$g_{\pi NN}^2/4\pi$                &14.3      &14.3      & 14.3 & 14.3       \\
$g_{\pi NN}^{(0)^2}/4\pi$              &5.77    &5.49    & 6.82    &6.08     \\
$g_{\sigma NN}g_{\sigma \pi\pi}^{(s)}/4\pi$    &-0.49  &-0.50   & -0.39   &-0.39     \\
$g_{\sigma NN}g_{\sigma \pi\pi}^{(v)}/4\pi$    &-1.52 &-1.40   &-1.43  &-1.40     \\
$g_{\rho NN}g_{\rho\pi\pi}/4\pi$           &3.05    &2.87    &2.68   &2.90     \\
$\delta$                       &0.65    &1.06   &1.26   &1.10 \\
$g_{\pi N\Delta}^2/4\pi$               &0.29   &0.29  &0.33    &0.34      \\
$g_{\pi N\Delta}^{(0)^2}/4\pi$             &0.17   &0.17   &0.18   &0.18      \\
$Z$                        &-0.13 &-0.075&-0.065 &-0.029   \\
$\kappa_V^\rho$                    &1.45   &1.19 &1.41  &1.55     \\
\\
$\Lambda_N$                    &1300   &1321  &1373    &1239     \\
$\Lambda_\sigma$                   & 653  & 681   & 400    &648      \\
$\Lambda_\rho$                     &1431  &1637   &2272   &1548     \\
$\Lambda_\Delta$                   &1522  &1542  &1507    &1429     \\
$\Lambda_\pi$                      &682     &666    &767    &859       \\
\\
\\
\hline
\end{tabular}
\end{small}
\end{center}
\caption{The parameters of the constructed meson-exchange
   models, defined by Eqs. (29)-(36), are compared.
   The models are constructed by using the three dimensional
   reduction schemes of Thompson (T2,T10) and Kadyshevsky (K2,K10).
   T2 (T10) and K2 (K10) are models with n=2 (n=10)
   in defining the form factor of Eq. (\ref{ff}).}
\end{table}

\newpage

\begin{figure}[tbp]
\begin{center}
\epsfig{file=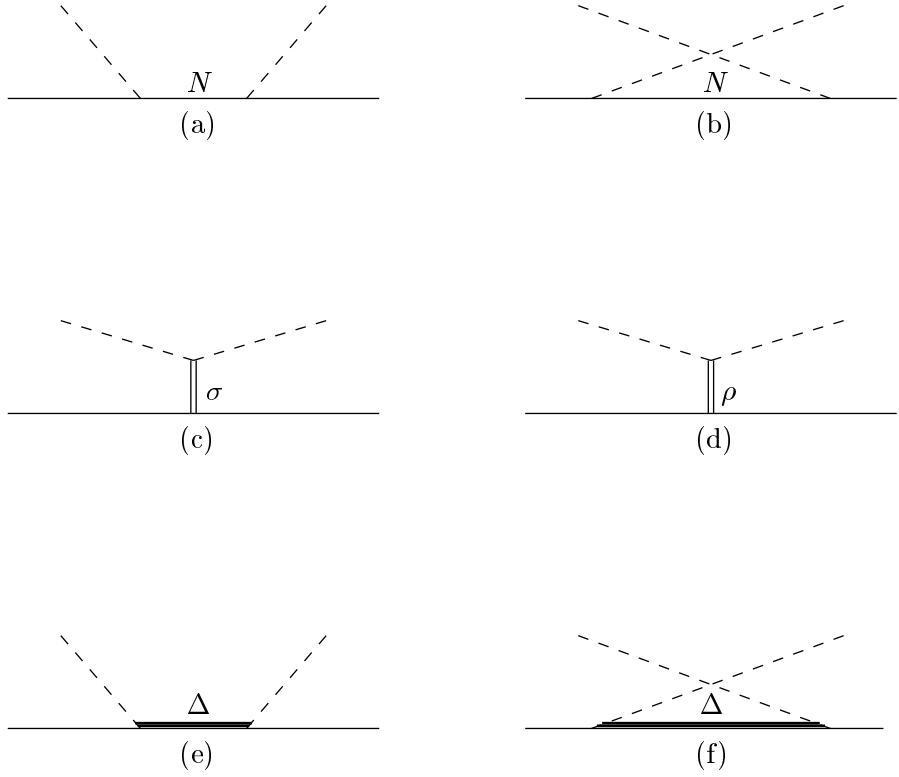,bbllx=50,bblly=400,bburx=500,bbury=600}
\end{center}
\caption{The driving terms of our model. (a) direct Born term, (b) u-channel nucleon
exchange, (c) t-channel $\sigma$ exchange, (d) t-channel $\rho$ exchange, (e) s-channel
$\Delta$ excitation, and (f) u-channel $\Delta$ exchange.}\label{fig1}
\end{figure}
\begin{figure}[tbp]
\begin{center}
\epsfig{file=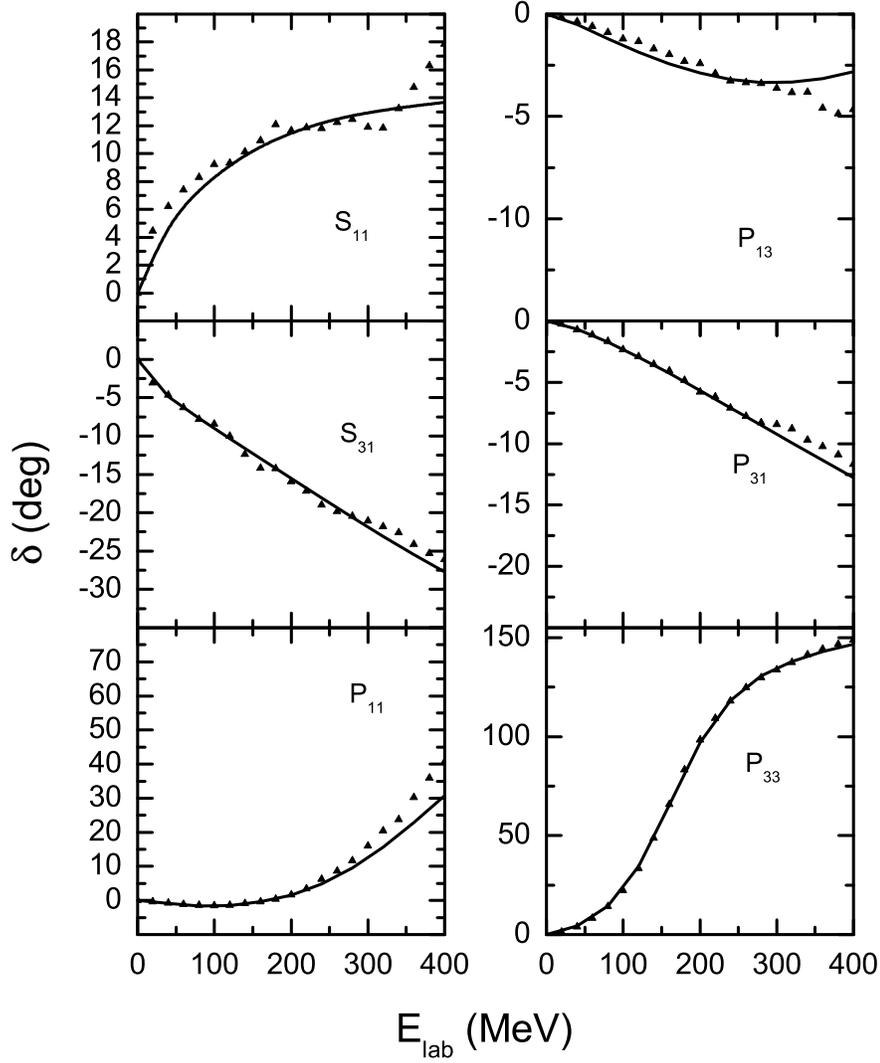,width=5.1in}
\end{center}
\caption{Our model predictions for $\pi N$ phase shifts in S- and P-waves obtained
within Kadyshevsky reduction scheme and with the use of a $n=2$  form factor of Eq.
(\ref{ff}). The data (solid triangles) are from \cite{VPI97}.}\label{fig2}
\end{figure}
\begin{figure}[tbp]
\begin{center}
\epsfig{file=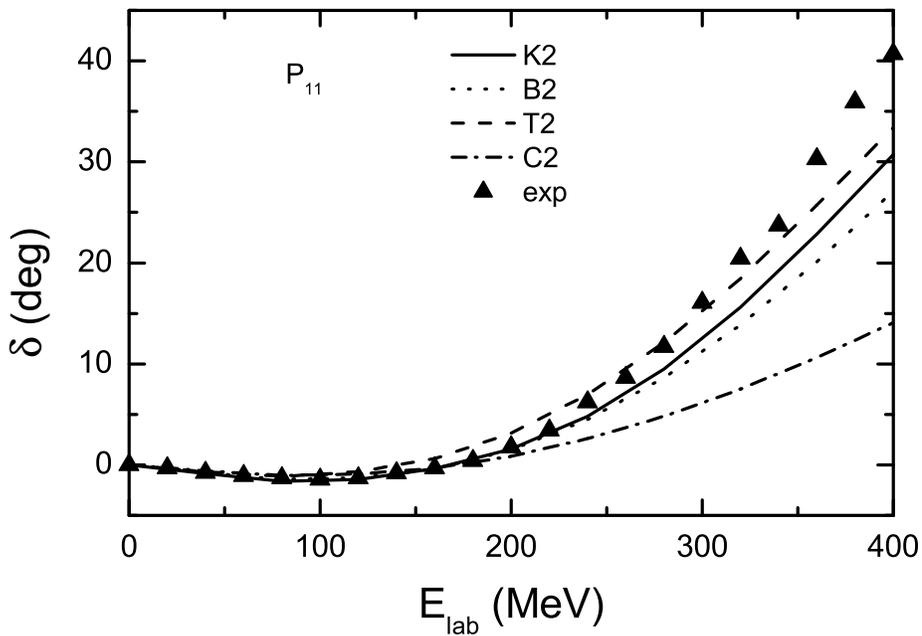,width=5in}
\end{center}
\caption{Our model predictions for $P_{11}$ phase shifts obtained within Kadyshevsky
(K2), Blankenbecler-Sugar (B2), Thompson (T2), and Cooper-Jennings (C2) reduction
schemes, all with $n=2$ form factor. Data (solid triangles) are from Ref. \cite{VPI97}.  }\label{fig3}
\end{figure}
\begin{figure}[tbp]
\begin{center}
\epsfig{file=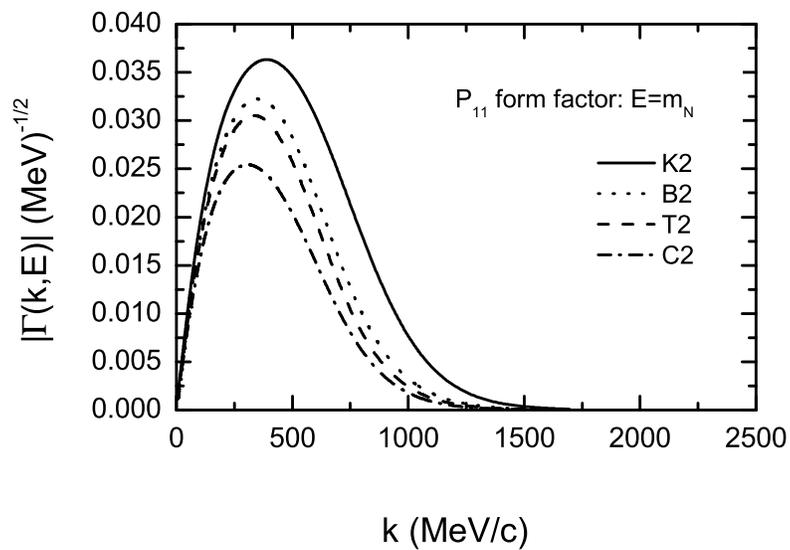,width=5in}
\end{center}
\caption{Our model predictions for the dressed $\pi NN$ vertex function
$\Gamma(k,E)$, obtained with various reduction schemes and $n=2$ form factor.}\label{fig4}
\end{figure}
\begin{figure}[tbp]
\begin{center}
\epsfig{file=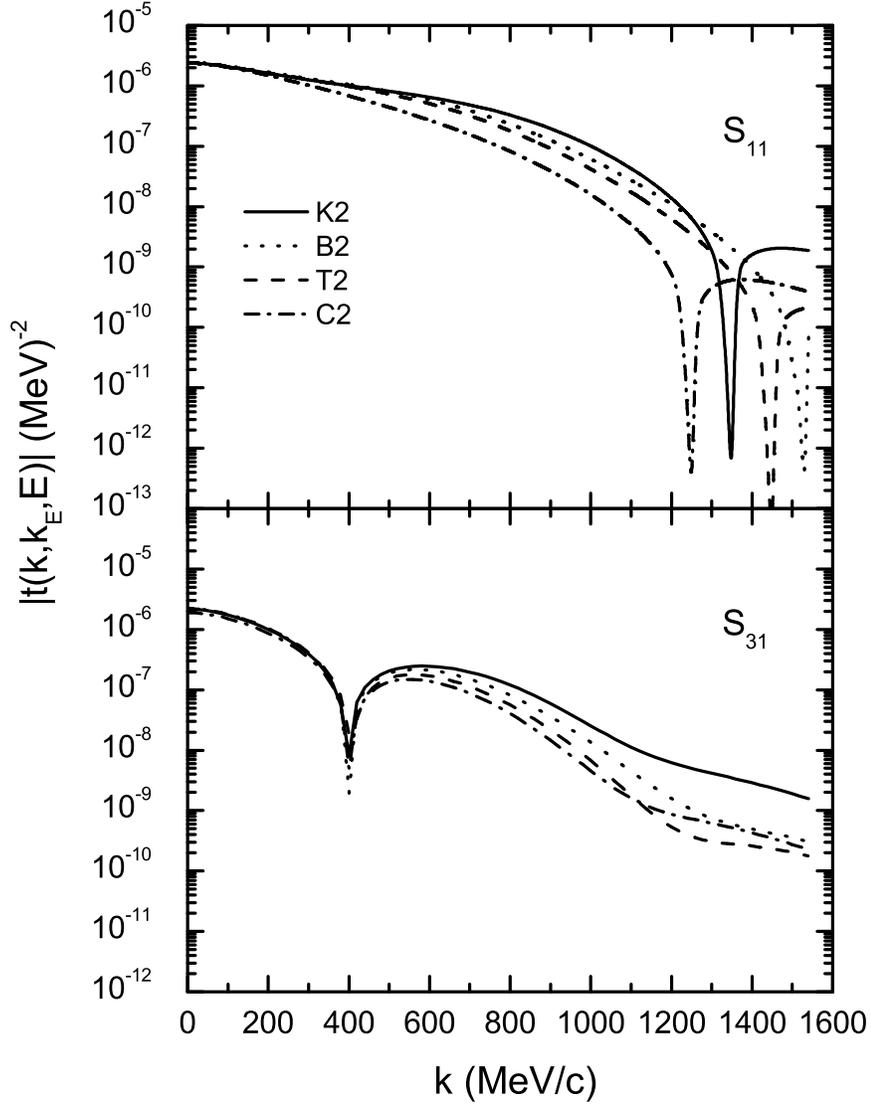,width=5in}
\end{center}
\caption{Our model predictions for the half-off-shell $t-$matrix elements in $S_{11}$
and $S_{31}$ channels at pion lab. energy $1 \, MeV$ above threshold, obtained with four different reduction
schemes and $n=2$ form factor.}\label{fig5}
\end{figure}
\begin{figure}[tbp]
\begin{center}
\epsfig{file=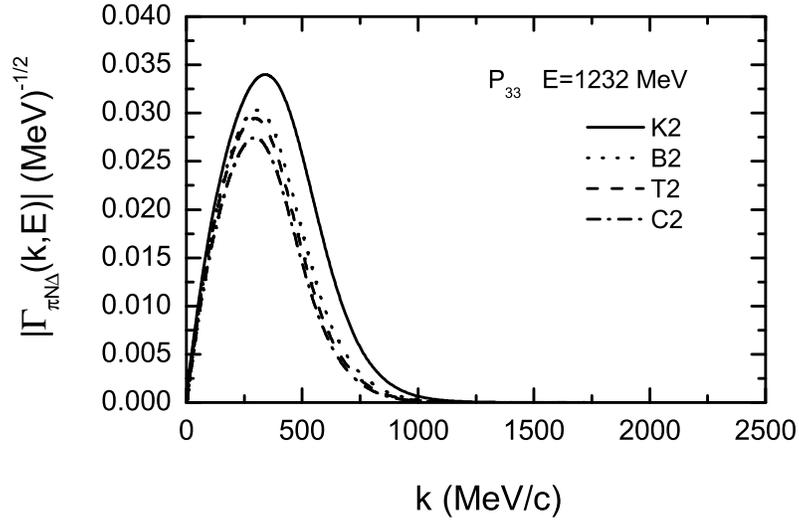,width=5in}
\end{center}
\caption{Our model predictions for the dressed $\pi N\Delta$ vertex $\Gamma_{\pi
N\Delta}$ at $E=1232 MeV$, obtained with various reduction schemes and $n=2$ form
factor.}\label{fig6}
\end{figure}
\begin{figure}[tbp]
\begin{center}
\epsfig{file=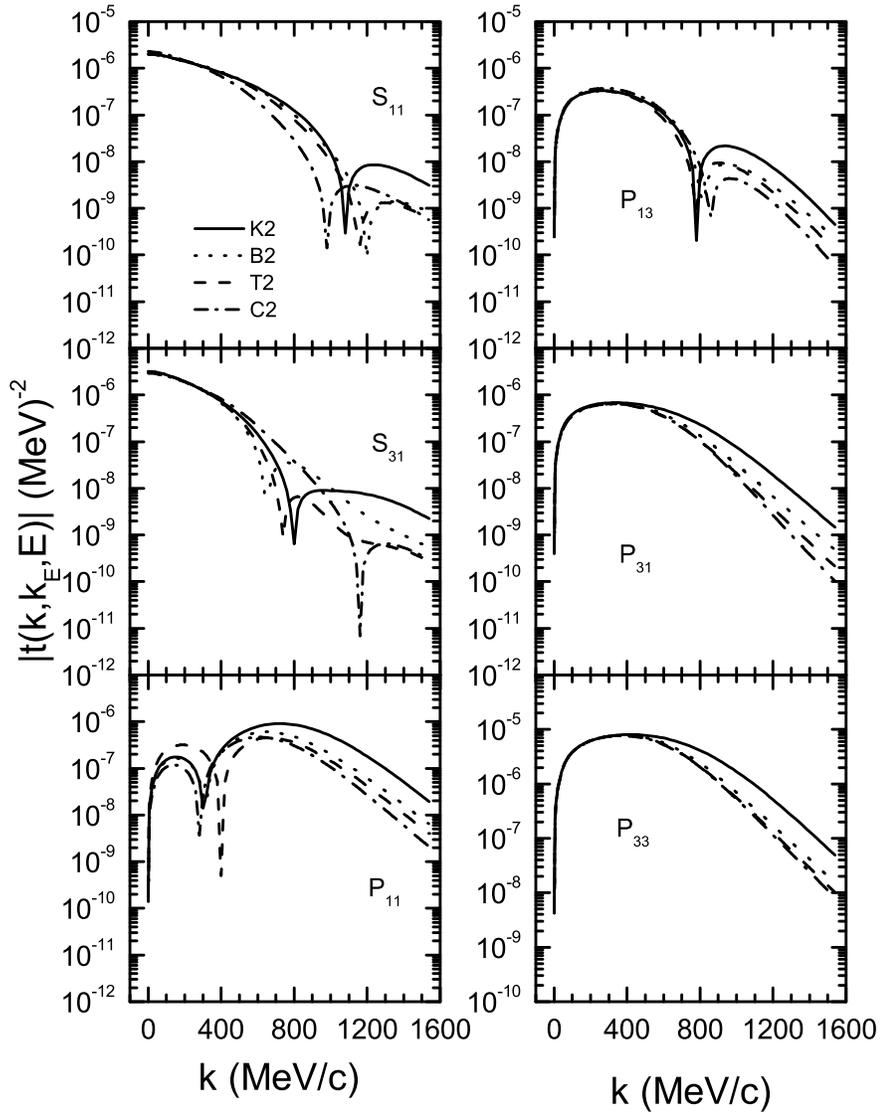,width=5in}
\end{center}
\caption{Same as Fig. 2 but at $E=1232 \,MeV$ and for all S- and P-waves.}\label{fig7}
\end{figure}
\end{document}